# Multiple Temporal Compression as a method to control the generation of bright and dark solitons

André C. A. Siqueira [1,*], Palacios G. [1], Mario B. Monteiro [1], Albert S. Reyna [2], Boris A. Malomed [3,4], Edilson L. Falcão-Filho [1], and Cid B. de Araújo [1].

[1] *Departamento de Física, Universidade Federal de Pernambuco,*
*50670-901 Recife-PE, Brazil.*
[2] *Programa de Pós-Graduação em Engenharia Física, Unidade Acadêmica do Cabo de Santo Agostinho, Universidade Federal Rural de Pernambuco, Cabo de Santo Agostinho, Pernambuco 54518-430, Brazil.*
[3] *Department of Physical Electronics, School of Electrical Engineering, Faculty of Engineering,*
*Tel Aviv University, Tel Aviv 69978, Israel.*
[4] *Instituto de Alta Investigación, Universidad de Tarapacá, Casilla 7D, Arica, Chile.*

*Corresponding author: andrechaves.physics@gmail.com
To be published in Phys. Rev. A.*

## Abstract

Recently published works have shown that the Multiple Temporal Compression (MTC) method is a more efficient approach for generation of multiple bright solitons in stacked waveguides, in comparison to the traditional soliton-fission technique. In the present paper, we performed systematic computer simulations of the appropriate generalized nonlinear Schrödinger equation to extend the MTC method for generation of dark solitons through controlled interactions of bright solitons in waveguiding systems with the normal group-velocity dispersion. Further, the generated dark solitons are demonstrated to be useful for accurately governing the dynamics of bright solitons generation in complex systems based on stacked waveguides. Therefore, we report scenarios that provide mitigation or switching of the energy transfer during bright-soliton collisions, along with the ability of the cascaded MTC processes to increase the number of the generated solitons. These results underscore the potential of the MTC as a versatile method for managing the propagation dynamics of multiple temporal bright and dark solitons produced by a single input pulse, with possible applications for the design of optical devices.

## I) INTRODUCTION

The use of temporal bright and dark solitons in various photonic based technologies, including the operation of ultrafast pulsed lasers, optical communications and data processing, laser sensing, manipulations of quantum information, etc., is a subject that has been attracting many experimental and theoretical works [1–12]. The solitons are maintained by the balance between the phase changes induced by the linear group-velocity dispersion (GVD), characterized by the respective coefficient ($\beta_2$), and the nonlinear self-phase modulation (SPM), determined by the nonlinear refractive index ($n_2$) [12–20]. Thus, the temporal bright and dark solitons exist at $\beta_2 n_2 < 0$ and $\beta_2 n_2 > 0$, respectively [19].

Based on these conditions, the creation of bright solitons was initially proposed in self-focusing ($n_2 > 0$) optical fibers operating in the anomalous dispersion regime [20]. According to the nonlinear Schrödinger equation (NLSE), which is the commonly known model for the dynamics of the optical wave in the fiber, bright solitons do not appear in the normal-GVD regime, as the combined phase chirp, induced by GVD and SPM, results in the temporal broadening of the pulse. Nonetheless, various strategies have been developed to exploit nonlinear optical phenomena in favor of the production of robust temporal solitons carried by wavelengths that feature normal GVD. Particularly, in studies using metal nanoparticle composites [21–30] and birefringent crystals [31–37], a negative cascaded-quadratic nonlinearity was employed to compensate the intrinsic positive cubic nonlinearity, leading to the observation of bright solitons when $\beta_2 > 0$.

Generating dark solitons in the normal-GVD regime is straightforward, as the NLSE admits currently known solutions of this type [4, 12, 19]. This was also experimentally tested by copropagating two delayed pulses in a self-focusing and normal dispersion medium, which generated a sinusoidally modulated pulse due to temporal interference between the trailing edge (blue side) of the first pulse and the leading edge (red side) of the delayed one [38–48]. As the pulse propagates further, this temporal modulation leads to the generation of dark pulses with finite background (dark soliton-like patterns), asymptotically approaching the creation of the fundamental dark solitons [45–48].

Beyond the bright- and dark-soliton propagation in normal-GVD media, other strategies have also aimed to generate multiple solitons. One traditional approach involves fission of higher-order solitons, in the framework of which higher-order solitons may split into a set of fundamental (bright) solitons [49], which may also proceed in the regime of the Newton's cradle (NC) formed by the soliton set [50]. In optical fibers, the Raman effect, higher-order dispersion, and self-steepening affect the fission process [49–54].

As a new strategy enhancing the generation of multiple bright solitons from a single input pulse propagating in normal-GVD media, we have recently introduced the Multiple Temporal Compression (MTC) method. It generates bright soliton pairs, starting from the edges to the center of the initial pulse, launched into a stacked system composed of self-focusing layers followed by self-defocusing ones (cf. a similar setting in the spatial domain, proposed earlier in Ref. [55]). Compared to the conventional higher-order soliton fission, the MTC method has the potential to double the number of bright solitons generated from the same input pulse by facilitating the energy redistribution and promoting configurations favorable to soliton collisions, including manifestations of the NC phenomenology [56, 57]. Furthermore, bright solitons generated by the MTC method exhibit high robustness, as they are formed at different temporal positions in the parent pulse. In contrast to that, solitons formed by the fission of higher-order solitons tend to concentrate around the pulse center, making them more susceptible to perturbations due to the interaction with additional components produced by the soliton fission (dispersive waves) [58].

In this paper, we extend the capabilities of the MTC method, introduced in Refs. [57, 58], to develop a nonlinearity-management procedure [59] for governing the generation of bright and dark temporal solitons. Thus, by controlling the interaction between solitons, we demonstrate the capability to transform bright solitons into dark soliton-like pulses, and vice versa, as well as to multiply the number of temporal solitons generated by the cascaded MTC process.

## II) THE THEORETICAL CONSIDERATION

The generation and transmission of temporal solitons by dint of the MTC method is addressed considering the propagation of light pulses in the stacked one-dimensional waveguide system composed of lossless segments with opposite signs of the nonlinearity. In principle, losses are introduced by reflection of light at junctions between the segments. However, the reflection, caused by the difference in the nonlinearity coefficients between the segments, rather than a discontinuity of the refractive index, is a weak effect. If necessary, the reflection loss may be compensated by additional linear amplification of the light pulses.

The nonlinear light propagation in this setting is governed by the generalized nonlinear Schrödinger equation (GNLSE) [19]:

$$\frac{\partial A(z,T)}{\partial z} - \left(\sum_{n\geq 2}^{\infty} \beta_n \frac{i^{n+1}}{n!}\frac{\partial^n A}{\partial T^n}\right) = i\gamma_0\left(1 + \frac{i}{\omega_0}\frac{\partial}{\partial T}\right)\left((1-f_R)A|A|^2 + f_R A \int_0^{\infty} h_R(\tau)|A(z,T-\tau)|^2 d\tau\right), \quad (1)$$

where $A = A(z,T)$ is the slowly varying envelope amplitude of the electric field, $\beta_n$ ($n \geq 2$) are the GVD coefficients, and $\gamma_0 = \omega_0 n_2(\omega_0)/cA_{\text{eff}}$ is the nonlinearity parameter, with $n_2(\omega_0)$ being the nonlinear refractive index at the central frequency $\omega_0$, while $A_{\text{eff}}$ and $c$ stand for the effective modal area and speed of light in vacuum, respectively. In line with Ref. [57], we consider the waveguides based on undoped fused silica with parameters $\beta_2 = +3.63 \times 10^{-2}$ ps$^2$m$^{-1}$, $\beta_3 = +2.75 \times 10^{-5}$ ps$^3$m$^{-1}$, $\beta_4 = -1.10 \times 10^{-8}$ ps$^4$m$^{-1}$, $\beta_5 = +3.15 \times 10^{-11}$ ps$^5$m$^{-1}$, $\beta_6 = -8.00 \times 10^{-14}$ ps$^6$m$^{-1}$, $\beta_7 = +2.50 \times 10^{-16}$ ps$^7$m$^{-1}$, and $\gamma_0 = 2.5$ W$^{-1}$km$^{-1}$, corresponding to the carrier wavelength 800 nm [57,60]. In addition, the full dependence of the silica dispersion coefficients on the wavelength is shown in Fig. 1. In the case of self-defocusing, we take $\gamma_0 = -2.5$ W$^{-1}$km$^{-1}$, which occurs in fused silica fibers doped by silver nanoparticles [57]. Actually, this choice is just an example, as the MTC technique is not limited to the assumption of having a fixed value of the GVD coefficient and fixed absolute values of the nonlinearity coefficient over the spectrum. Other materials with normal GVD and controllable nonlinearity, such as birefringent crystals [31–37], can also be considered to explore the effects of the nonlinearity management in this context. In fact, the key requirement for implementing the MTC technique is the use of media with opposite signs of the nonlinearity in the given dispersion range.

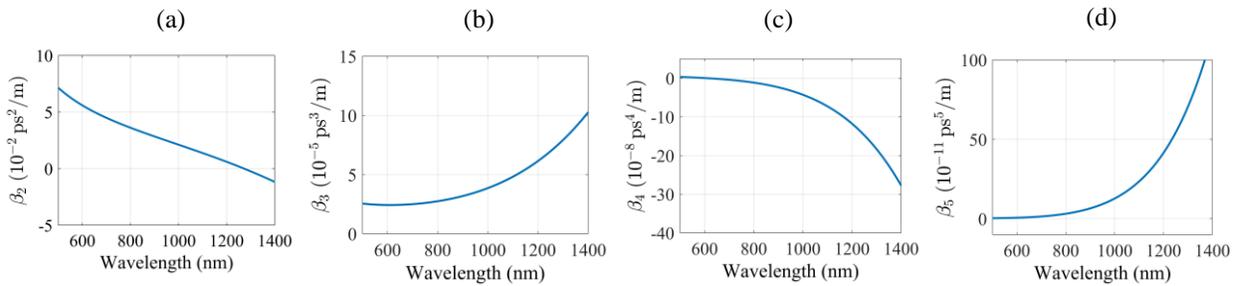

Figure 1: Silica dispersion coefficients as a function of wavelength: (a) $\beta_2$, (b) $\beta_3$, (c) $\beta_4$, and (d) $\beta_5$.

The right-hand side of Eq. (1) represents the cubic nonlinear terms, viz., the self-phase modulation (SPM), self-steepening and intrapulse Raman scattering (IRS). To model the latter one, we set $f_R = 0.18$, which is a fractional contribution of the delayed Raman response to the nonlinear polarization, and employ the Raman response function

hR($\tau$) = $\tau_1(\tau_1^{-2} + \tau_2^{-2})$ exp $(-\tau/\tau_2)$ sin$(\tau/\tau_1)$, with $\tau_1$ = 12.2 fs, and $\tau_2$ = 32.0 fs, which are typical parameters of fused silica [19].

The fourth-order Runge-Kutta algorithm in the interaction picture (RK4IP) was used to simulate the GNLSE [61]. Both bright and dark temporal solitons were produced, using a Gaussian input, $A(0,T) = \sqrt{P_0}$exp $[-0.5(1.665T/T_{FWHM})^2]$, with a central wavelength $\lambda_0$ = 800 nm, temporal width $T_{FWHM}$ = 90 fs, and two values of peak power, $P_0$ = 70 kW and 420 kW. The nonlinearity of the medium (self-focusing or self-defocusing) determines the generation of dark or bright temporal solitons.

## III) RESULTS AND DISCUSSION

### a) The generation of bright solitons by the MTC method

Bright temporal solitons were generated by dint of the MTC method applied to the two-stacked waveguide system in the normal-GVD regime, where the first and second waveguides exhibit the self-focusing and defocusing nonlinearity, respectively. Thus, as the pulse propagates along the first waveguide (of length $L_1$ = 10 mm), it accumulates a positive nonlinear chirp under the action of SPM ($n_2 > 0$) and normal GVD ($\beta_2 > 0$), as shown in the spectrograms displayed in Figs. 2(a) and 2(b), corresponding to the input peak powers of 70 kW and 420 kW, respectively. Then, this positively chirped pulse enters the second waveguide, with $n_2 < 0$ and length ($L_2$) of up to 290 mm, where partial compensation of the nonlinear phase occurred at the beginning of $L_2$.

Unlike the generation of bright solitons from zero-chirp input pulses, the main idea behind the use of the nonlinear positive chirp accumulated in the first waveguide is to reshape the pulse to generate, in the second waveguide, pairs of temporal solitons that are symmetric with respect to the pulse center which are responsible for the implementation of the MTC technique as described in Refs. [57, 58].

In other words, throughout the first medium ($\beta_2 n_2 > 0$), the accumulation of the nonlinear positive chirp produces, due to the normal dispersion, pulse broadening where the red components are located at the leading edge of the pulse, while the blue components are at the trailing edge. As this pulse with the positive nonlinear chirp propagates in the second medium ($\beta_2 n_2 < 0$), SPM reverses the direction of frequency generation, leading to the creation of blue components at the front of the pulse (in the region that already had red components) and red components at the back of the pulse (in the region that already had blue components). As a result, instead of the pulse compressing toward its center (the conventional higher-order-soliton fission [58]), multiple bright solitons are generated through the MTC process along the pulse temporal profile.

Figs. 2(c,e) and 2(d,f) are examples of two and eight bright solitons, along with the corresponding spectrograms, generated for input peak powers of 70 kW and 420 kW, respectively. It is noteworthy that, in the absence of the first waveguide, which corresponds of the usual scheme of the higher-order-soliton fission, only one and five bright solitons are observed for the same input powers. The difference underscores the efficiency of the MTC method.

Figures 2(c,d) reveal the acceleration experienced by the fundamental solitons upon their generation, which is caused by the soliton self-frequency shift (SSFS). This observation is corroborated by analyzing the evolution of the central wavelengths of both the trailing-edge (TE) and leading-edge (LE) solitons, as shown in Fig. 3(a). Note that, for the input pulse power of 70 kW, during the initial propagation through the length of 120 mm, both the TE and LE solitons exhibit a pronounced redshift caused by IRS, in the normal-GVD regime. Therefore, given the relationship of the Raman-induced frequency shift for the fundamental bright solitons, $\Omega_p(z) \propto -z/T_{\text{FWHM}}^4$ [19], it is apparent that the TE soliton, which exhibits a shorter temporal width than its LE counterpart ($T_{\text{FWHM}}^{\text{TE}} = 18$ fs and $T_{\text{FWHM}}^{\text{LE}} = 25$ fs, at L = 90 mm), experiences greater acceleration due to SSFS. Consequently, collisions between the fundamental solitons are likely to occur. As illustrated in Fig. 2(c), at the peak power of 70 kW, after propagating approximately 120 mm, the TE soliton inelastically collides with the LE soliton. This partially out-of-phase soliton collision (with the phase difference $\Delta\phi = 0.6\pi$) results in spectral broadening, which is characteristic for the repulsive interaction [62], as shown in Fig. 3(b). Consequently, a significant weakening (strengthening) of the SSFS due to the loss (gain) of energy by the TE (LE) soliton for distances beyond 120 mm is observed, as illustrated by Fig. 3(c), where the evolution of the peak power and temporal width is shown. After the collision, the TE and LE solitons acquire approximately $T_{\text{FWHM}}^{\text{TE}} = 82$ fs, $P_0^{\text{TE}} = 16.5$ kW and $T_{\text{FWHM}}^{\text{LE}} = 15$ fs, $P_0^{\text{LE}} = 177.6$ kW at $L = 300$ mm, respectively.

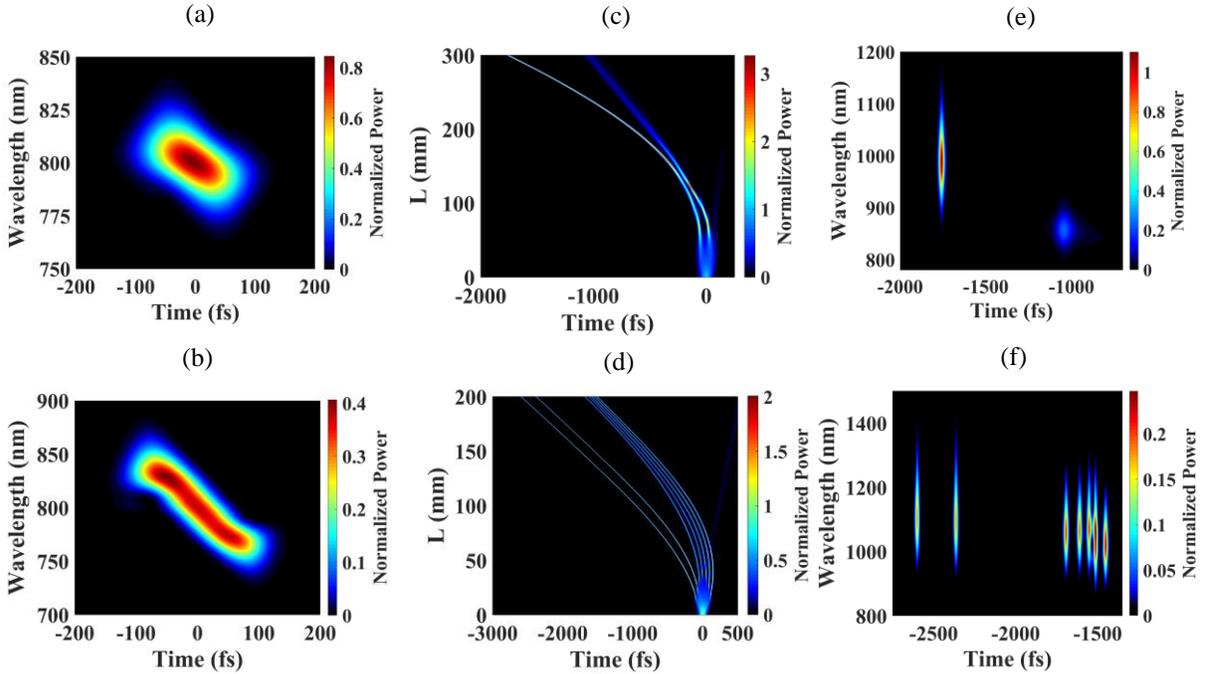

Figure 2: The generation of bright solitons by means of the MTC method, for input peak power of 70 kW (a,c,e), and 420 kW (b,d,f). Panels (a,b) and (e,f) display spectrograms in the output plane of the first waveguide (with length $L_1 = 10$ mm) and second waveguide (with $L_2 = 290$ mm in (e) and $L_2 = 190$ m in (f)), respectively. Spatiotemporal trajectories of the solitons are shown in (c,d). Different total propagation distances are employed solely for the presentation purposes.

A still more interesting phenomenon is observed when the incident pulse carries a power which is high enough to generate multiple bright solitons. Notably, for the input power of 420 kW (Fig. 2(d)), out of the eight bright solitons generated, seven exhibit a regular spacing in time, with similar peak powers and time durations ($\approx$ 10 fs and $\approx$ 355 kW at L = 200 mm), suggesting comparable accelerations via SSFS. Notably, this behavior persists despite a slight increase in the central wavelengths of the solitons, from TE to LE, attributed to the accumulation of the positive chirp in the $L_1$ waveguide (Figs. 2(a,b)).

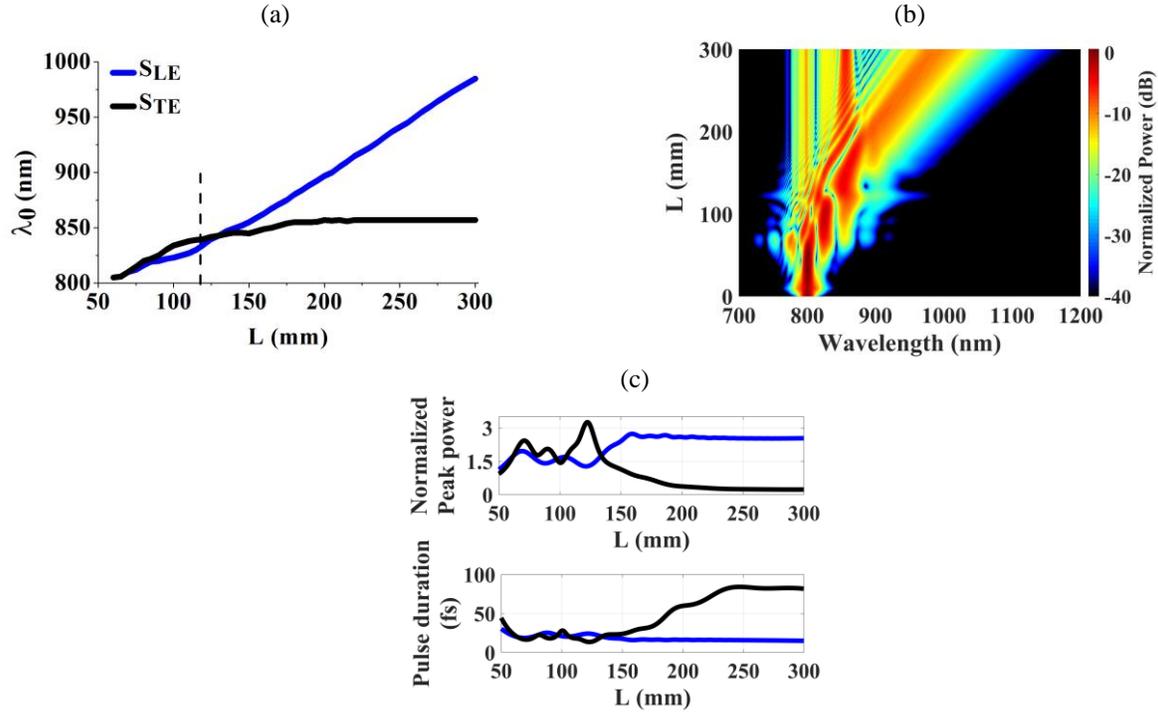

Figure 3: (a) The evolution of the central wavelengths of the LE (leading-edge, $S_{LE}$) and TE (trailing-edge, $S_{TE}$) solitons, respectively, for the 70 kW input pulse, in the configuration shown in Fig. 2(c). The dashed vertical line denotes the position of the collision between the solitons. (b) The spectral evolution during the soliton generation and interaction driven by the MTC method, accompanied by their respective evolution of the peak power and pulse duration (FWHM) (c). In (a) and (c), the blue and black curves represent $S_{LE}$ and $S_{TE}$, respectively.

The eighth bright soliton, which is generated near the central region of the input pulse, exhibiting a lower peak power and larger temporal width, experiences a delay, which is produced by positive third-order dispersion, $\beta_3$ [19]. Consequently, it undergoes inelastic collisions with solitons emerging at the rear of the pulse, being eventually ejected from the soliton chain. This series of soliton interactions resembles the above-mentioned optical-soliton version of NC, as previously reported in Ref. [50]. At this point, it is worth mentioning that, in the configurations explored in this work, the self-steepening effect does not play a significant role in the temporal dynamics, with the IRS being the effect chiefly affecting the soliton dynamics.

## b) The generation of dark solitons by MTC-mediated bright-soliton collisions

Temporal dark solitons can also be efficiently generated by means of the MTC method, leveraging the well-established mechanism of the interference mechanism between two delayed pulses in a waveguide with $\gamma_0 \beta_2 > 0$ [38–48]. One strategy to achieve this condition is inserting a third waveguide, exhibiting similar self-focusing characteristics as the first one, just before the collision of the bright solitons. For example, as depicted in Fig. 2(c), the collision between two bright solitons occurs, approximately, at the propagation distance 120 mm after the incident pulse has propagated, with the input power of 70 kW. Accordingly, the third waveguide with $\gamma_0 = 2.5$ W$^{-1}$km$^{-1}$ is introduced at position $L = 115$ mm. Thus, the stacked-waveguide system comprises the first self-focusing waveguide ($\beta_2 n_2 > 0$) of length $L_1 = 10$ mm, followed by the second self-defocusing one ($\beta_2 n_2 < 0$) of length $L_2 = 105$ mm, and finally the third self-focusing segment ($\beta_2 n_2 > 0$) of length $L_3 = 85$ mm, totaling 200 mm in length.

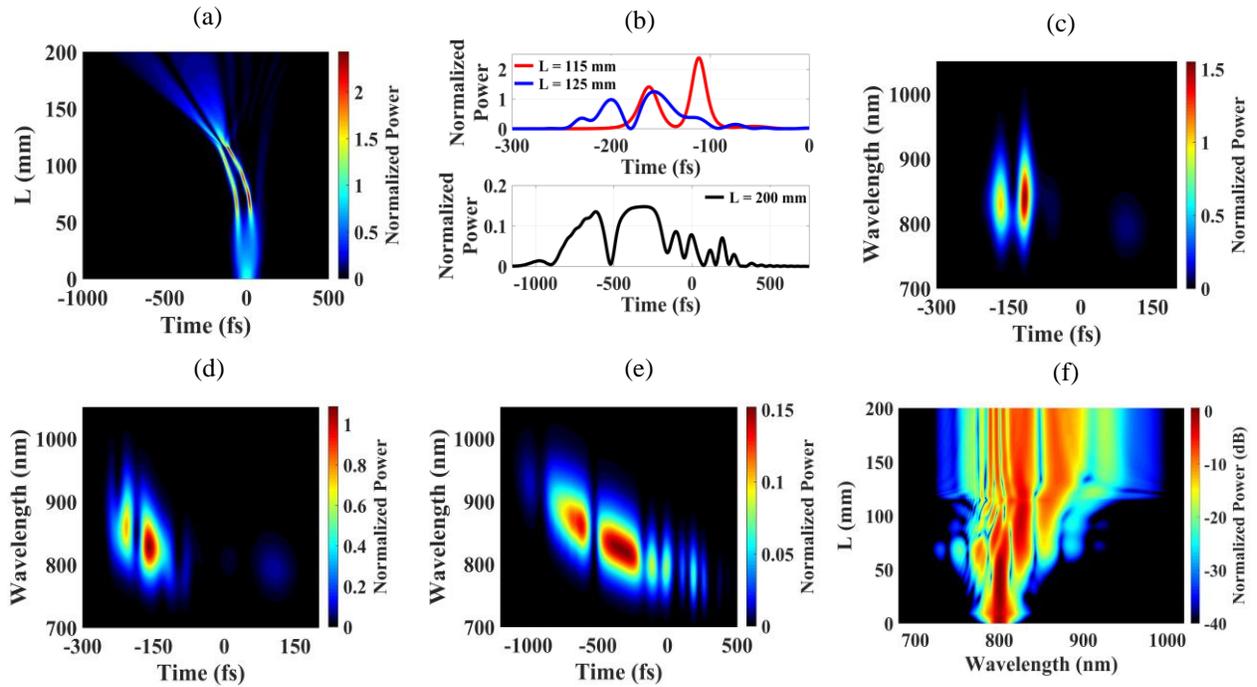

Figure 4: a) The temporal evolution leading to the generation of dark solitons by means of MTC-mediated collisions between bright solitons in the stacked three-waveguide system. (b) Temporal profiles of the pulse at the exit of the second waveguide (L = 115 mm, the red curve), subsequently passing 10 mm (L = 125 mm, the blue curve) and 85 mm (L = 200 mm, the black curve) in the third waveguide. The respective spectrograms are calculated for the following positions: (c) L = 115 mm (showing two bright solitons), (d) L = 125 mm (initiating the formation of a dark soliton-like pulse), and (e) L = 200 mm (showcasing a dark soliton-like pulse). (f) The spectral evolution of the dark-soliton generation by means of the MTC-mediated bright-soliton collisions.

Figure 4(a) illustrates the evolution of the propagating pulse in this system, showcasing the interferometric pattern resulting from the interaction between two delayed pulses at $L \gtrsim 115$ mm, which leads to the generation of a dark pulse that evolves into a dark soliton with a finite background. This behavior was corroborated by analyzing the temporal profiles of the pulse and their respective spectrograms in the course of the propagation, as shown in Figs. 4(b) and 4(c-e), respectively. For example, the red curve in Fig. 4(b), corresponding to the propagation distance $L$ = 115 mm in the stacked-waveguide system, features the temporal profile of two bright solitons, separated by a 50−fs temporal delay, heading towards collision. The spectrogram in Fig. 4(c) reveals that the LE soliton has power $P_{LE}$ = 98.5 kW, temporal width $T_{FWHM}^{LE}$ = 23 fs, and the central wavelength $\lambda_{LE}$ = 828 nm, while the TE soliton exhibits $P_{TE}$ = 166.5 kW, $T_{FWHM}^{TE}$ = 17.5 fs, and $\lambda_{TE}$ = 837 nm. However, with the addition of the third self-focusing waveguide, the collision is arrested, and, instead, a dark pulse is formed between the two bright ones, with a positive nonlinear chirp (as indicated by the blue curve in Fig. 4(b) and the spectrogram in Fig. 4(d)). Finally, the black curve in Fig. 4(b) and the spectrogram in Fig. 4(e) demonstrate passing the distance of 85 mm by this dark pulse in the third waveguide, until it reaches L = 200 mm, maintaining a nearly constant temporal width ($T_{FWHM}$ = 70 fs) due to the spreading out of neighboring ultrashort pulses with slightly different central wavelengths. Note that the spectral evolution of the pulse, presented in Fig. 4(f), reveals that the spectral broadening observed as the result of the soliton collision (Fig. 3(b)) persists due to the structure of the dark soliton, ranging between 723 nm and 1000 nm at −40 dB for L $\gtrsim$ 130 mm. Hence, akin to the results reported in Ref. [45], the dark-soliton-like pulses generated by the MTC-mediated bright-soliton collisions are not the fundamental dark solitons (with the infinite background) in the strict sense, but they exhibit similar behavior.

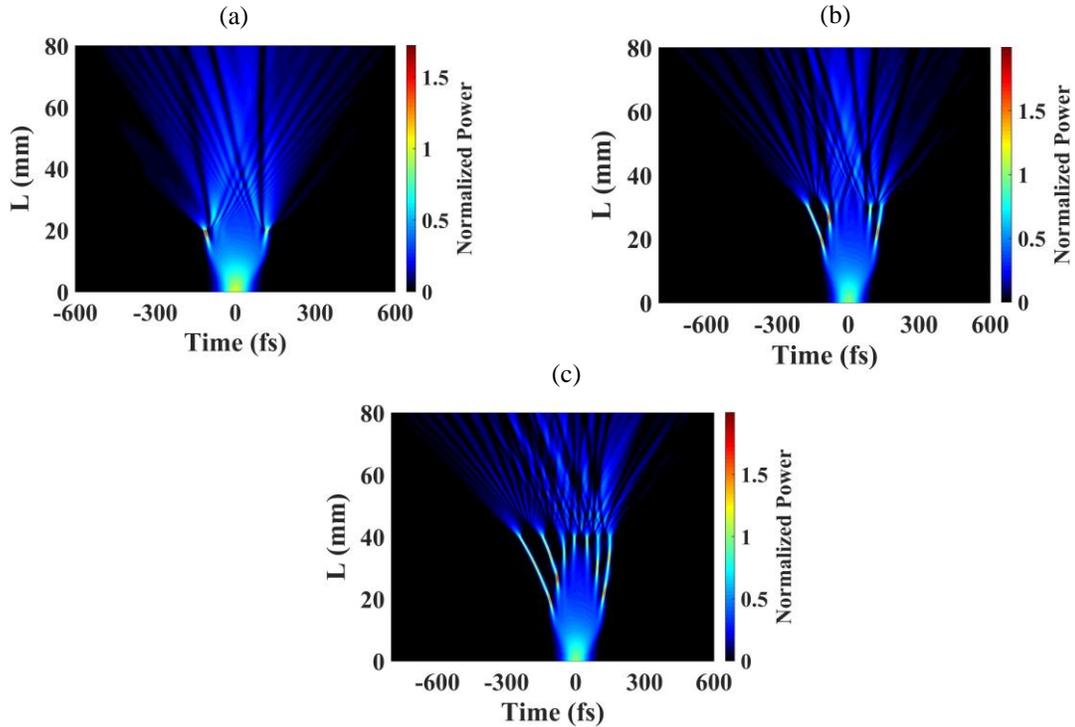

Figure 5: The generation of multiple dark solitons through the interference of bright solitons by dint of the MTC method. In all cases, the three-stacked waveguide system is used, with the first waveguide length fixed as $L_1$ = 10 mm, while the second and third ones vary: (a) $L_2$ = 10 mm and $L_3$ = 60 mm, (b) $L_2$ = 20 mm and $L_3$ = 50 mm, and (c) $L_2$ = 30 mm and $L_3$ = 40 mm. The peak power of the input pulse is 420 kW.

While two bright solitons were employed to generate a dark soliton, the interference between multiple bright solitons with counterpropagating frequencies can produce multiple dark solitons. For demonstration purpose, we used an input pulse power of 420 kW, which is sufficient to generate up to eight bright solitons (see Fig. 2(d)). Thus, the generation of a pair of dark solitons was achieved by exploring a small propagation distance in the second waveguide ($L_2 = 10$ mm) and incorporating the third one with length $L_3 = 60$ mm, as shown in Fig. 5(a). Along its propagation path, gray solitons are observed as a consequence of multiple interference. By increasing the propagation distance in the second waveguide, a larger number of dark-soliton pairs are generated, facilitated by a larger number of bright solitons used as the seeds. This is evidenced in Figs. 5(b) and 5(c) for $L_2 = 20$ mm and 30 mm, respectively. As the number of the dark solitons increases, collisions between them become observable too.

## c) Controlling the propagation dynamics of temporal solitons by means of the MTC method

The nonlinearity management provided by the MTC method offers an innovative approach to control the temporal soliton dynamics, with potential applications for the use in ultrafast optical devices. For instance, generating a dark soliton not only prevents collisions between bright solitons, as shown in Fig. 4(a), but also enables the reversal of the energy transfer direction between adjacent pulses, adding a fourth self-focusing waveguide to the MTC setup. This approach was demonstrated via the numerical simulation of a four-stacked-waveguide system, totaling the propagation length of 400 mm, with $L_1 = 10$ mm ($\beta_2 n_2 > 0$), $L_2 = 105$ mm ($\beta_2 n_2 < 0$), $L_3 = 10$ mm ($\beta_2 n_2 > 0$), and $L_4 = 275$ mm ($\beta_2 n_2 < 0$). Note that the fourth (third) waveguide has the same dispersion coefficients and nonlinear parameters as the second (first) one.

Figure 6(a) illustrates how introducing the fourth waveguide, following a brief propagation in the third self-defocusing one (between $L = 115$ mm and 125 mm), reverts the TE and LE pulses back into bright solitons. Notably, the analysis of the peak-power evolution in the course of the propagation (see Fig. 6(b)) reveals that the TE soliton reaches a peak power $\approx 7.7$ times higher than its LE counterpart, contrasting with the solitons' behavior observed in Fig. 2(c) and 3(c). As a consequence, the TE (LE) soliton undergoes a more pronounced (negligible) redshift via IRS, according to the spectral evolution shown in Fig. 6(c). This phenomenon is best observed by analyzing the evolution of the central wavelength $\lambda_0$, as shown in Figure 6(d). Note that in the region corresponding to the third waveguide, where the dark soliton emerges, blue components are transferred from the LE pulse to the TE one, inducing a redshift (blueshift) in the LE (TE) pulse. Subsequently, in the fourth waveguide (between $L = 125$ mm and $L = 400$ mm), where bright solitons reappear, $\lambda_0 = 875$ nm of the LE soliton remains constant, while $\lambda_0$ of the TE one continues to increase under the action of IRS. Additionally, interaction between both bright solitons is observed at $L \approx 310$ mm, resulting from their disparate accelerations. Comparing the energy exchange between the bright solitons at $L = 250$ mm (prior to the collision) and $L = 400$ mm (post-collision), almost negligible energy exchange between the solitons indicates an elastic collision, in an approximately in-phase form (with the phase difference $\Delta\phi = 0.056\pi$), making it difficult to identify which pulse corresponds to what one in the pre-collision set [62]. Therefore, examining the results presented in Figs. 3(a,c) and Fig. 6(a,b), it is evident that the nonlinearity management provided by the MTC method can switch the energy transfer between the solitons and the occurrence of the soliton redshift induced by IRS.

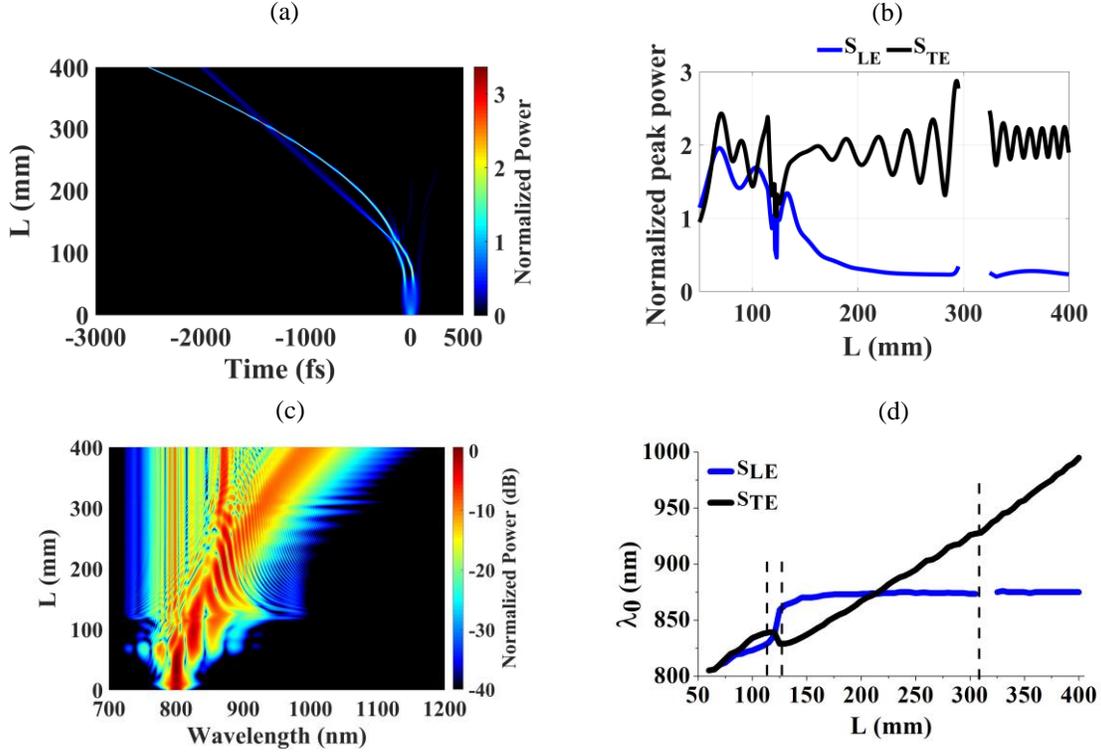

Figure 6: (a) Temporal, (b) peak-power and (c) spectral evolution of the bright soliton generation by means of the MTC method, using the four-stacked waveguide with lengths $L_1 = 10$ mm, $L_2 = 105$ mm (bright solitons), $L_3 = 10$ mm (the dark soliton) and $L_4 = 275$ mm (bright solitons). (d) The central wavelengths of $S_{LE}$ and $S_{TE}$ (cf. Fig. 3) as functions of the propagation distance. The region between the first two vertical dashed lines denotes the propagation in the third waveguide (where the dark soliton emerges), while the third vertical dashed line indicates the position of the attractive collision between the bright solitons.

In another case, using the four-stacked waveguide system, but with an extended propagation distance in the third waveguide ($L_3 = 30$ mm), it is possible to realize a situation in which both the TE and LE pulses, responsible for the generation of the dark solitons, accumulate a positive nonlinear chirp sufficient to initiate a cascaded MTC process, resulting in the generation of new multiple bright solitons. As seen in Fig. 7(a), after propagating the distance of 145 mm ($L_1 = 10$ mm, $L_2 = 105$ mm, $L_3 = 30$ mm), the TE pulse generates two bright solitons in the fourth waveguide, while the LE pulse behaves as a third bright soliton. Moreover, tripling the nonlinearity of the fourth waveguide to $3\gamma_0 = -7.5$ W$^{-1}$km$^{-1}$ amplifies the cascaded MTC process, yielding five bright solitons, with two originating from the LE pulse and three from the TE one. These solitons undergo out-of-phase (repulsive) collisions, as shown in Fig. 7(b). These collisions may be regulated by adjusting the fourth waveguide's nonlinearity. Note that, for the nonlinearity of $6\gamma_0 = -15$ W$^{-1}$km$^{-1}$, a configuration less susceptible to collisions between adjacent bright solitons is observed, showcasing solitons with a regular temporal spacing (see Fig. 7(c)). These results underscore the potential of the MTC method for the creation of temporal solitons that can be used for the design of temporal optical demultiplexing devices.

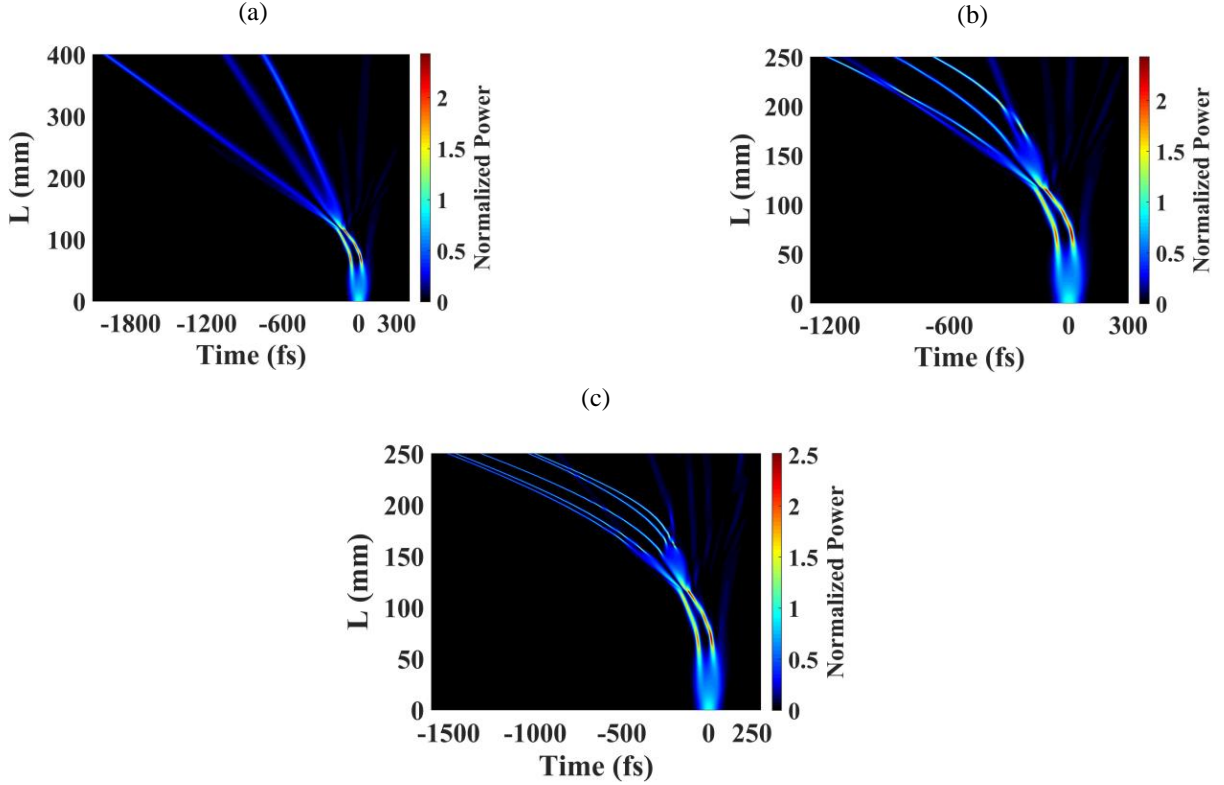

Figure 7: The multiple-bright-soliton generation by the cascaded MTC process in the four-stacked waveguide system. In all scenarios, the propagation distances are fixed as $L_1 = 10$ mm, $L_2 = 105$ mm, and $L_3 = 30$ mm. The fourth waveguide is inserted after the propagation of distance $L_1 + L_2 + L_3 = 145$ mm, with the nonlinearity of (a) $\gamma_0 = -2.5$ $W^{-1}km^{-1}$, (b) $3\gamma_0$, and (c) $6\gamma_0$.

## IV) Conclusions

The results reported in this paper underscore the remarkable capability of the MTC (Multiple Temporal Compression) method to control the propagation dynamics of multiple bright and dark solitons, generated from a single input pulse, in normal-GVD waveguiding systems. By means of simulations of the respective generalized nonlinear Schrödinger equation, we have explored systems composed of up to four stacked waveguides, featuring alternating self-focusing and defocusing nonlinearities. Employing the set of two stacked waveguides — one self-focusing and one defocusing — we have observed the formation of accelerated pairs of bright solitons, whose numbers increase with the intensity of the input pulse. By manipulating their accelerations and thus inducing collisions between soliton pairs, we have reported the generation of dark pulses arising from the interference between adjacent bright ones. Introducing the third self-focusing waveguide facilitates the propagation of dark pulses which maintain their shapes, thus giving rise to dark soliton-like pulses with a finite background. Notably, the increase in the number of bright solitons leads to the generation of multiple dark solitons through the interference processes. Moreover, our numerical results reveal strategies for mitigating inelastic collisions between bright solitons, and for reverting the direction of the energy transfer between them, in the course of the generation of dark solitons.

These techniques hold significant promise for the design of ultrafast optical devices. The MTC method also demonstrates its versatility by allowing the transformation of dark solitons back into bright ones, incorporating the fourth self-defocusing waveguide into the setup. This transformative capability can be further harnessed through cascading MTC processes to increase the number of the produced bright solitons, thus further extending the method's applicability for various purposes. Potential materials suitable for the soliton generation by means of the MTC technique are: (i) transparent $\chi^{(2)}$ nonlinear bulk crystals, where the nonlinearity with $n_2 < 0$ is achieved via the cascaded $\chi^{(2)}$ process; (ii) composite nanocrystallite-doped materials, where the effective optical nonlinearity may be influenced by quantum-mechanical and dielectric-confinement effects; and (iii) periodically poled fibers or ferromagnetic crystals that emulate negative n2 through the cascaded $\chi^{(2)}$ interactions. These possibilities suggest the experimental implementation of the soliton generation by the MTC technique.

## Acknowledgments

We acknowledge the financial support of the Brazilian Agencies: Conselho Nacional de Desenvolvimento Científico e Tecnológico (CNPq, Grant: 406441/2023-5), the Coordenação de Aperfeiçoamento de Pessoal de Nível Superior (CAPES), and the Fundação de Amparo à Ciência e Tecnologia do Estado de Pernambuco (FACEPE). The postdoctoral fellowship of A.C.A.S. and P.G. were provided by Office of Naval Research (ONR), research Grant of ONR global: N62909-23-1-2014. The work of B.A.M. was supported, in part, by the Israel Science Foundation, through grant No. 1695/22.

## References

[1] L. F. Mollenauer, and K. Smith, *Demonstration of soliton transmission over more than 4000 km in fiber with loss periodically compensated by Raman gain*, Opt. Lett. 13, 675 (1988).

[2] F. Abdullaev, S. Darmanyan, and P. Khabibullaev, *Optical Solitons* (Springer-Verlag, New York, 1993).

[3] Y. S. Kivshar, and B. Luther-Davies, *Dark optical solitons: physics and applications*, Phys. Rep. 298, 81 (1998).

[4] Y. S. Kivshar and G. P. Agrawal, *Optical Solitons: From Fibers to Photonic Crystals* (Academic Press, San Diego, 2003).

[5] A. Blanco-Redondo, C. Husko, D. Eades, Y. Zhang, J. Li, T. F. Krauss, and B. J. Eggleton, *Observation of soliton compression in silicon photonic crystals*, Nat. Commun. 5, 3160 (2014).

[6] D. E. Chang, V. Vuletić, and M. D. Lukin, *Quantum nonlinear optics—photon by photon*, Nat. Photonics 8, 685 (2014).

[7] L. Salmela, N. Tsipinakis, A. Foi, C. Billet, J. M. Dudley, and G. Genty, *Predicting ultrafast nonlinear dynamics in fibre optics with a recurrent neural network*, Nat. Mach. Intell. 3, 344 (2021).

[8] S. Boscolo, J. M. Dudley, and C. Finot, *Predicting nonlinear reshaping of periodic signals in optical fibre with a neural network*, Opt. Commun. 542, 129563 (2023).

[9] M. Mabed, F. Meng, L. Salmela, C. Finot, G. Genty, and J. M. Dudley, *Machine learning analysis of instabilities in noise-like pulse lasers*, Opt. Express 30, 15060 (2022).

[10] L. Salmela, C. Lapre, J. M. Dudley, and G. Genty, Machine learning analysis of rogue solitons in supercontinuum generation, Sci. Rep. 10, 9596 (2020).

[11] B. A. Malomed, *Multidimensional solitons*. (AIP Publishing: Melville, NY, 2022).

[12] S. Yang, Q. Y. Zhang, Z. W. Zhu, Y. Y. Qi, P. Yin, Y. Q. Ge, L. Li, L. Jin, L. Zhang, and H. Zhang, *Recent advances and challenges on dark solitons in fiber lasers*, Opt. Laser Technol. 152, 108116 (2022).

[13] J. P. Gordon, *Theory of the soliton self-frequency shift*, Opt. Lett. 11, 662 (1986).

[14] F. M. Mitschke and L. F. Mollenauer, *Discovery of the soliton self-frequency shift*, Opt. Lett. 11, 659 (1986).

[15] K. Tai, A. Hasegawa, and N. Bekki, *Fission of optical solitons induced by stimulated Raman effect*, Opt. Lett. 13, 392 (1988).

[16] A. V. Husakou, and J. Herrmann, *Supercontinuum generation of higher-order solitons by fission in photonic crystal fibers*, Phys. Rev. Lett. 87, 203901 (2001).

[17] A. Demircan, and U. Bandelow, *Analysis of the interplay between soliton fission and modulation instability in supercontinuum generation*, Appl. Phys. B: Lasers Opt. 86, 31 (2007).

[18] A. A. Sysoliatin, A. K. Senatorov, A. I. Konyukhov, L. A. Melnikov, and V. A. Stasyuk, *Soliton fission management by dispersion oscillating fiber*, Opt. Express 15, 16302 (2007).

[19] G. P. Agrawal, *Nonlinear Fiber Optics*, 5th ed. (Academic Press, Oxford, 2013).

[20] A. Hasegawa, and F. Tappert, *Transmission of stationary nonlinear optical pulses in dispersive dielectric fibers. I. Anomalous dispersion*, Appl. Phys. Lett. 23, 142-144 (1973).

[21] N. Zhavoronkov, R. Driben, B. A. Bregadiolli, M. Nalin, and B. A. Malomed, *Observation of asymmetric spectrum broadening induced by silver nanoparticles in a heavy-metal oxide glass*, Europhys. Lett. 94, 37011 (2011).

[22] A. S. Reyna and C. B. de Araújo, *High-order optical nonlinearities in plasmonic nanocomposites—a review*, Adv. Opt. Photonics 9, 720 (2017).

[23] *Metal Nanostructures for Photonics*. Edited by L. R. P. Kassab and C. B. de Araújo. Nanophotonics Series (Elsevier Ltd., Cambridge, MA, 2019) ISBN:978-0–08-102378-5.

[24] A. S. Reyna and C. B. de Araújo, *Beyond third-order optical nonlinearities in liquid suspensions of metal-nanoparticles and metal-nanoclusters*, J. Opt. 24, 104006 (2022).

[25] F. R. Arteaga-Sierra, A. Antikainen, and G. P. Agrawal, *Soliton dynamics in photonic-crystal fibers with frequency-dependent Kerr nonlinearity*, Phys. Rev. A 98, 013830 (2018).

[26] S. Bose, A. Sahoo, R. Chattopadhyay, S. Roy, S. K. Bhadra, and G. P. Agrawal, *Implications of a zero-nonlinearity wavelength in photonic crystal fibers doped with silver nanoparticles*, Phys. Rev. A 94, 043835 (2016).

[27] R. Driben and J. Herrmann, *Solitary pulse propagation and soliton-induced supercontinuum generation in silica glasses containing silver nanoparticles*, Opt. Lett. 35, 2529 (2010).

[28] S. Zhao, R. Guo, and Y. Zeng, *Effects of frequency-dependent Kerr nonlinearity on higher-order soliton evolution in a photonic crystal fiber with one zero-dispersion wavelength*, Phys. Rev. A 106, 033516 (2022).

[29] S. Bose, R. Chattopadhyay, S. Roy, and S. K. Bhadra, *Study of nonlinear dynamics in silver-nanoparticle-doped photonic crystal fiber*, J. Opt. Soc. Am. B 33, 1014 (2016).

[30] S. Bose, R. Chattopadhyay, and S. K. Bhadra, *Dispersive shock mediated resonant radiations in defocused nonlinear medium*, Opt. Commun. 412, 226 (2018).

[31] K. Beckwitt, F. W. Wise, L. Qian, L. A. Walker, and E. Canto-Said, *Compensation for self-focusing by use of cascade quadratic nonlinearity*, Opt. Lett. 26, 1696 (2001).

[32] H. Guo, X. Zeng, B. Zhou, and M. Bache, *Few-cycle solitons and supercontinuum generation with cascaded quadratic nonlinearities in unpoled lithium niobate ridge waveguides*, Opt. Lett. 39, 1105 (2014).

[33] R. DeSalvo, D. J. Hagan, M. Sheik-Bahae, G. Stegeman, E. W. Van Stryland, and H. Vanherzeele, *Self-focusing and self-defocusing by cascaded second-order effects in KTP*, Opt. Lett. 17, 28 (1992).

[34] S. Ashihara, J. Nishina, T. Shimura, and K. Kuroda, *Soliton compression of femtosecond pulses in quadratic media*, J. Opt. Soc. Am. B 19, 2505 (2002).

[35] M. Bache, O. Bang, W. Krolikowski, J. Moses, and F. W. Wise, *Limits to compression with cascaded quadratic soliton compressors*, Opt. Express 16, 3273 (2008).

[36] R. Šuminas, G. Tamošauskas, V. Jukna, A. Couairon, and A. Dubietis, *Second-order cascading-assisted filamentation and controllable supercontinuum generation in birefringent crystals*, Opt. Express 25, 6746 (2017).

[37] M. Conforti and F. Baronio, *Extreme high-intensity and ultrabroadband interactions in anisotropic $\beta-BaB_2O_4$ crystals*, J. Opt. Soc. Am. B 30, 1041 (2013).

[38] J. E. Rothenberg, J. *Dark soliton trains formed by visible pulse collisions in optical fibers*, Opt. Commun. 82, 107 (1991).

[39] S. A. Gredeskul, and Y. S. Kivshar, *Dark-soliton generation in optical fibers*, Opt. Lett. 14, 1281 (1989).

[40] J. E. Rothenberg, and H. K. Heinrich, *Observation of the formation of dark-soliton trains in optical fibers*, Opt. Lett. 17, 261 (1992).

[41] W. Zhao, and E. Bourkoff, *Generation, propagation, and amplification of dark solitons*, J. Opt. Soc. Am. B 9, 1134 (1992).

[42] J. A. R. Williams, K. M. Allen, N. J. Doran, and P. Emplit, *The generation of quasi-continuous trains of dark soliton-like pulses*, Opt. Commun. 112, 333 (1994).

[43] A. M. Weiner, J. P. Heritage, R. J. Hawkins, R. N. Thurston, E. M. Kirschner, D. E. Leaird, and W. J. Tomlinson, *Experimental observation of the fundamental dark soliton in optical fibers*, Phys. Rev. Lett. 61, 2445 (1988).

[44] W. J. Tomlinson, R. J. Hawkins, A. M. Weiner, J. P. Heritage, and R. N. Thurston, *Dark optical solitons with finite-width background pulses*, J. Opt. Soc. Am. B 6, 329 (1989).

[45] T. Marest, C. M. Arabí, M. Conforti, A. Mussot, C. Milián, D. V. Skryabin, and A. Kudlinski, *Emission of dispersive waves from a train of dark solitons in optical fibers*, Opt. Lett. 41, 2454 (2016).

[46] S. Zhao, H. Yang, Y. Huang, and Y. Xiao, *Generation of tunable ultra-short pulse sequences in a quasi-discrete spectral supercontinuum by dark solitons*, Opt. Express 27, 23539 (2019).

[47] C. Milián, T. Marest, A. Kudlinski, and D. V. Skryabin, *Spectral wings of the fiber supercontinuum and the dark-bright soliton interaction*, Opt. Express 25, 10494 (2017).

[48] Y. Chen, H. Yang, J. Rong, and H. Hu, *Effects of the induced seed pulse on generation of ultra-short pulse sequences by dark solitons*, Opt. Commun. 500, 127318 (2021).

[49] J. M. Dudley, G. Genty, and S. Coen, *Supercontinuum generation in photonic crystal fiber*, Rev. Mod. Phys. 78, 1135 (2006).

[50] R. Driben, B. A. Malomed, A. V. Yulin, and D. V. Skryabin, *Newton's cradles in optics: From N-soliton fission to soliton chains*, Phys. Rev. A 87, 063808 (2013).

[51] J. M. Dudley, and S. Coen, *Numerical simulations and coherence properties of supercontinuum generation in photonic crystal and tapered optical fibers*, IEEE J. Sel. Top. Quantum Electron. 8, 651 (2002).

[52] J. K. Ranka, R. S. Windeler, and A. J. Stentz, *Visible continuum generation in air–silica microstructure optical fibers with anomalous dispersion at 800 nm*, Opt. Lett. 25, 25-27 (2000).

[53] C. Ciret, S. Gorza, C. Husko, G. Roelkens, B. Kuyken, and F. Leo, *Physical origin of higher-order soliton fission in nanophotonic semiconductor waveguides*, Sci. Rep. 8, 17177 (2018).

[54] J. Herrmann, U. Griebner, N. Zhavoronkov, A. Husakou, D. Nickel, J. C. Knight, W. J. Wadsworth, P. St. J. Russell, and G. Korn, *Experimental evidence for supercontinuum generation by fission of higher-order solitons in photonic fibers*, Phys. Rev. Lett. 88, 173901 (2002).

[55] I. Towers and B. A. Malomed, *Stable (2+1)-dimensional solitons in a layered medium with sign-alternating Kerr nonlinearity*, J. Opt. Soc. Am. B 19, 537 (2002).

[56] A. C. A. Siqueira, B. A. Malomed, C. B. de Araújo, and E. L. Falcão-Filho, *Generation of multiple ultrashort solitons in heterogeneous medium with self-focusing and defocusing nonlineari*ties, in Proceedings of the Latin America Optics And Photonics Conference, (Optica Publishing Group, 2022), pp. M4A-3 (2022).

[57] A. C. A. Siqueira, E. L. Falcão-Filho, B. A. Malomed, and C. B. de Araújo, *Generation of multiple ultrashort temporal solitons in a third-order nonlinear composite medium with self-focusing and self-defocusing nonlinearities*, Phys. Rev. A 107, 063519 (2023).

[58] A. C. A. Siqueira, G. Palacios, A. S. Reyna, B. A. Malomed, E. L. Falcão-Filho, and C. B. de Araújo, *Generation of robust temporal soliton trains by the multiple-temporal-compression (MTC) method*, Opt. Commun. 545, 129723 (2023).

[59] B. A. Malomed, *Soliton Management in Periodic Systems* (Springer, New York, 2006).

[60] I. H. Malitson, *Interspecimen Comparison of the Refractive Index of Fused Silica*, J. Opt. Soc. Am. 55, 1205 (1965).

[61] J. Hult, *A Fourth-Order Runge–Kutta in the Interaction Picture Method for Simulating Supercontinuum Generation in Optical Fibers*, J. Lightwave Technol. 25, 3770 (2007).

[62] A. Antikainen, M. Erkintalo, J. Dudley, and G. Genty, *On the phase-dependent manifestation of optical rogue waves*, Nonlinearity. 25, R73 (2012).